\documentstyle{mn}
\begin{document}
\def\simlt{\mathrel{\rlap{\lower 3pt\hbox{$\sim$}}
        \raise 2.0pt\hbox{$<$}}}
\def\simgt{\mathrel{\rlap{\lower 3pt\hbox{$\sim$}}
        \raise 2.0pt\hbox{$>$}}}

\title[Variance and Skewness in the FIRST Survey]
{Variance and Skewness in the FIRST Survey}
\author[M. Magliocchetti, S.J.Maddox, O. Lahav, J.V.Wall]
{M.~Magliocchetti $^1$,
 S.J.Maddox $^1$, O.~Lahav $^1$, J.V.Wall $^2$\\
$^1$Institute of Astronomy, Madingley Road, Cambridge CB3 0HA\\
$^2$Royal Greenwich Observatory, Madingley Road, Cambridge CB3 0EZ}

\maketitle
\begin{abstract}

We investigate the large-scale clustering of radio sources in the 
FIRST 1.4-GHz survey by analysing the distribution function
({\it counts in cells}). 
We select a reliable sample from the the
FIRST catalogue, paying particular attention to the problem of how to
define single radio sources from the multiple components listed. 
We also consider the incompleteness of the catalogue.
We estimate the angular two-point correlation function $w(\theta)$, the
variance $\Psi_2$, and skewness $\Psi_3$ of the distribution for
the various sub-samples chosen on different criteria.
Both $w(\theta)$ and $\Psi_2$ show power-law behaviour with an amplitude 
corresponding a spatial correlation length of $r_0 \sim 10 h^{-1}$Mpc.
We detect significant skewness in the distribution, the first such
detection in radio surveys. This skewness is found to be
related to the variance through
$\Psi_3=S_3(\Psi_2)^{\alpha}$, with $\alpha=1.9\pm 0.1$, consistent with 
the non-linear gravitational growth of perturbations 
from primordial 
Gaussian initial conditions.
We show that the amplitude of variance and skewness are 
consistent with realistic models of galaxy clustering.
 
\end{abstract}

\begin{keywords}
galaxies: clustering - radio galaxies - large-scale structure
\end{keywords}

\section{INTRODUCTION}

Surveys of optical and infrared galaxies have revealed the rich
structure of the universe and have provided much information on the
large-scale structure out to redshifts $z\sim 0.1$ (e.g. APM,
Maddox et al., 1990 and IRAS, Fisher et al., 1993). In contrast
radio sources, although representing only a small fraction of all
galaxies, can be detected over significant cosmological distances (up
to $z\sim 4$), sampling much larger volumes of space and therefore with
the potential to provide information on much larger physical scales.

It had been suggested (Kaiser, 1984) that galaxies form preferentially
in high-density peaks of the underlying mass distribution. If it is
true, then the statistics of galaxy distributions provide us with
information, although biased, about the underlying  
matter distribution.

It has long been debated whether radio galaxies are clustered or
isotropic on the largest scale. The study by Webster
(1976), which looked at 8000 radio sources, found $<3\%$
variability in the number of sources in randomly-placed 1-Gpc
cubes. This led to a widely-accepted view that radio sources were
isotropically distributed. Even if this were not true, the large range
in intrinsic luminosity of radio sources might effectively wash out
structures when the distribution is projected onto the sky with
information about radial distribution effectively lost. Other studies
(Seldner \& Peebels (1981), Shaver \& Pierre (1989)) 
reported a detection of slight
clustering of nearby radio sources,
while Benn \& Wall (1995) and Baleisis et al. (1998) 
discussed other measures of anisotropy
in radio surveys. Clustering in the 4.85 Ghz Green Bank (87GB:
Gregory \& Condon, 1991) and in the Parkes-MIT-NRAO (PMN: Griffith \& Wright,
1993) catalogues was
studied by correlation-function analysis (Kooiman et al., 1995;
Sicotte, 1995; Loan, Wall \& Lahav, 1997). These studies indicated that
radio objects are actually more strongly clustered than local 
optically-selected
galaxies. This conclusion was confirmed by correlation analysis of
the FIRST survey (Cress et al., 1996).

One of the possible ways of investigating clustering properties of
radio sources is by means of the distribution function ({\it
counts in cells}) i.e. the probability of finding $N$ galaxies in a
cell of particular size and shape. This analysis includes all the
moments of the distribution function and therefore provides a more
complete description of large-scale structure. Furthermore it can be shown
(Peebles, 1980; Frieman \& Gazta\~naga, 1994) that the higher-order
moments of the galaxy distribution can be used as a test of non-linear
models for large-scale structure.

This paper presents a counts-in-cells analysis carried out for the
FIRST radio survey. We focus on the second and third moment of
the distribution together with the angular two-point correlation
function of the sample, and we test the predictions of different
cosmological models.  We report a detection of skewness
$S_3 =\Psi_3/\Psi_2^2=const$ 
(with $\Psi_2$ and $\Psi_3$ defined as the
second and the third moment of the distribution). Our measurement
accords with the hypothesis of non-linear growth of observed
structures by gravitational clustering from initially-Gaussian density
fluctuations.

In Section 2 we describe the catalogue in its original form and explain
the procedures providing us with modified samples
analysed in the rest of the paper. 
In Section 3 we present the results of our analysis
for the angular two-point correlation function, and 
the angular second and third moments of the distribution. Section 4
discusses the deprojection of our 2-d
measurements to estimate the quantities describing spatial distribution.
Section 5 summarises our conclusions.

\section{THE DATA}

\subsection{The Public Catalogue}
The FIRST (Faint Images of the Radio Sky at Twenty centimetres) survey
(Becker et al., 1995) began in the spring of 1993 and will eventually
cover 10,000 square degrees of the sky in the north Galactic cap.  
The VLA is being used in B-configuration to take 3-min snapshots of
23.5-arcmin fields arranged on a hexagonal grid.  The observations are
at 1.4~GHz and the beam-size is 5.4~arcsecs, with an rms sensitivity of
typically 0.14~mJy/beam. 
A map is produced for each field and sources are detected using an
elliptical Gaussian fitting procedure (White et al., 1997). The 5-rms
source detection limit is roughly 1~mJy. 
This survey is 50 times more sensitive than any previous large-area
radio survey (see White et al., 1997 for details), leading to a 
high surface density of objects in the catalogue ($\sim 100$ per square
degree). It therefore provides an excellent tool for investigating the
clustering properties of faint sources.

The catalogue is still in the process of construction, but
nevertheless is publically accessible.  We used the 27 Feb 1997
version which contains approximately 236,000 entries and is derived
from the 1993 through 1996 observations  covering about 2575
square degrees, including most of the area
$7^h20^m<{\rm RA}(2000)<17^h20^m$, $22.2^\circ<{\rm Dec}<42.5^\circ$. 
The sky coverage is available in the form of a map of rms
sensitivity at 3-arcmin resolution. 
To simplify our clustering analysis we restricted the area to 
$7^h44^m<{\rm RA}<17^h20^m$, $22.4^\circ<{\rm Dec}<41.8^\circ$,
which has essentially complete coverage. 

Within this area there are regions of low source-density because the
original catalogue includes only sources brighter than 5~rms.  In
calculating the counts-in-cells statistics in the following sections
we used the coverage map of the survey to `mask out' all those cells
in which the noise was large enough to reduce the number of images in the
catalogue, ie 0.2~mJy for a flux limit of 1~mJy, and 0.6~mJy for a flux
limit of 3~mJy.  When we analyse cells significantly larger than the
coverage pixels, we reject a cell only if the fraction of useful area
is less than 0.75. We keep cells with a ratio $\frac{useful
\;area}{total\;area}\;>\;0.75$, and correct the count in each
cell by the ratio of useful to total area.
For the angular correlation-function analysis we corrected for these
effects by generating random catalogues modulated by the rms
sensitivity from the coverage map.

The public catalogue is not immediately suitable for clustering analysis,
as there are spurious images from sidelobes of bright sources, and
single sources are frequently listed as multiple components.
These ``extra'' sources would artificially increase the observed
clustering in the catalogue.
Probable side-lobes have been identified using an oblique decision-tree
program and are flagged in the catalogue (White et al., 1997).  
We simply rejected all the flagged sources from the original catalogue.
We investigate the problem of multi-component sources in some
detail in the next section. 

We initially chose the flux limit of 1~mJy, corresponding to a
5-rms source detection (as discussed in more detail in Section
2.3). In this way we chose a sample of 189,689 sources from the raw
catalogue. We label this sample as C1. 

\subsection{Multi-component Sources} 
Radio sources often have widely-separated components
corresponding to the nucleus with hot-spots along and at the end of the jets.
If the shape of a source is complex, the source detection algorithm
fits several components to reproduce the shape.
These two effects mean that a single radio source can appear in the
catalogue split into two or more sub-components.
If analysed as single sources, these components would
seriously distort clustering measurements, with
effects particularly serious for higher order moments
which are more sensitive to tight groupings.

We have investigated  techniques to identify tight groups of sources
likely to be sub-components of a single radio source.  
For each group we replace all of the sub-components by a single source
with flux equal to the sum of the individual fluxes, placed at the
mean position. 
As described in the following sections, this
recombining procedure is of crucial importance in the analysis of the
clustering.

\begin{figure}
\vspace{8cm}  
\includegraphics{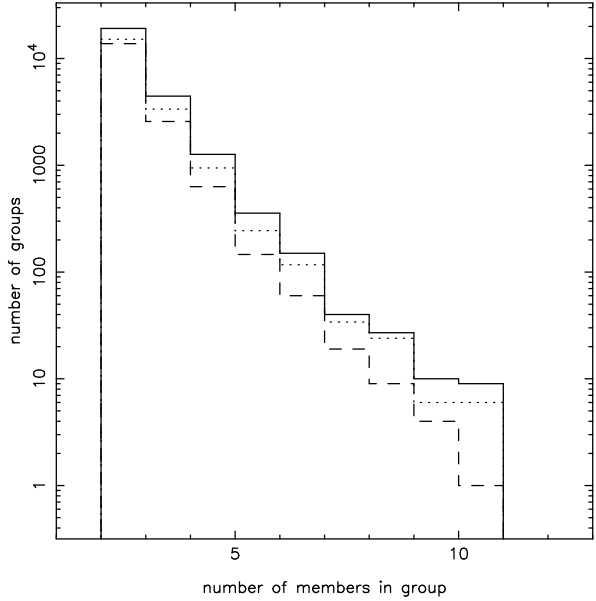} 
\caption{Number of component groups derived after the
  combining procedure vs the number of objects in each group. The
  solid line shows the results obtained by combining sources
  according to their separation ($\theta_{link}=0.02^\circ$); the
  dotted line is given by adding a procedure for removing
  spurious aggregations according to a flux-separation
  relationship, while the dashed line includes a constraint on the
  flux ratio.
\label{fig:grouphist} }
\end{figure}

Following Cress et al. (1996), using a percolation technique we first simply 
identified
all groups of sources within $0^.02^{\circ}$ of each other; each such group 
was replaced by a single source. 
This procedure found 25447 groups with a mean of 2.36 sources per
group. A histogram of the number of groups as a function of the number
of sources in the group is shown in Figure~\ref{fig:grouphist}. 
Since the position and flux of the new composite source is different
from the individual sub-components, it is possible that the new 
source can be grouped with other sources. 
To ensure that all groups were located, we re-ran the group-finding
procedure on the revised catalogue repeatedly until no new groups were 
found. This produced just 10 extra groups, leaving a catalogue of
155,084 sources. We refer to this sample as C2. 

\noindent

This procedure efficiently combines sub-components into 
a single composite source, but can also combine physically-distinct 
sources into a single source, with drastic effect on
the apparent clustering in the catalogue (see Figure~\ref{fig:wplot}). 
We have attempted to improve on this simple approach by varying the
link-length in the percolation procedure according to the flux of each
source. 
This is motivated by the well-known $\theta - S$ relation, extended by 
Oort (1987) to low flux densities, and shown to follow $\theta \propto 
\sqrt S$.

\begin{figure}
\vspace{8cm}  
\includegraphics{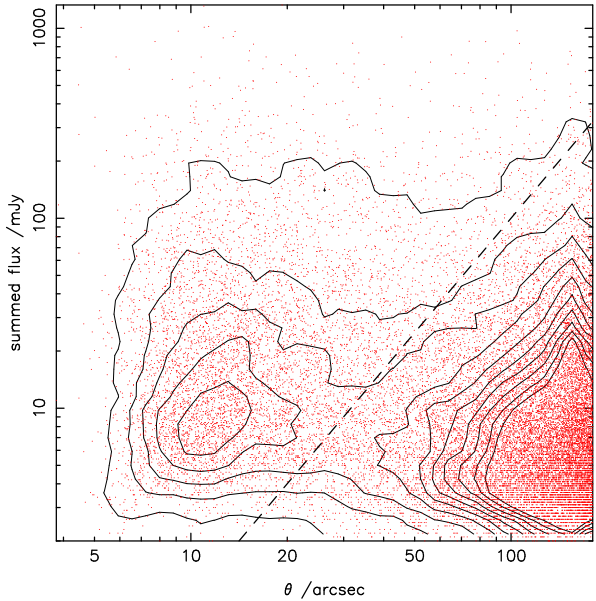} 
\caption{Sum of the fluxes of the components of double sources as a
  function of their separation. We chose to combine pairs which lie 
 to the left of the dashed line (see text for details).
\label{fig:pairsep} }
\end{figure}
\noindent

In order to determine a reasonable flux--separation relation, we plot the
sum of the fluxes of the components of each apparent double versus
their separation up to $ 0.05^\circ$, as shown in
Figure~\ref{fig:pairsep}.
Superimposed on this scatter plot, we  show contours of the
density of points in the $\theta$--flux plane.
There are two distinct regions of high density: one on the lower-right 
and the other on the
lower-left of the figure.
We inspected maps from the FIRST survey showing 4.5-arcmin squares
around a sample of these objects, and found that the peak on the left
consists mostly of sub-components of sources as evidenced by trailing 
substructures between them. The peak on the right consists of independent 
pairs of sources.

We set the maximum link-length to be proportional to the
square-root of the summed flux, $F_{TOT}$,
\begin{eqnarray}
\theta_{link}=100\left(\frac{F_{TOT}}{100}\right)^{0.5} arcsec.
\label{eqn:rlink} 
\end{eqnarray}
This line is shown by the
dashed line in Figure~\ref{fig:pairsep} and 
roughly follows the minimum source-density between the two peaks.
Varying the link-length with flux in this way 
combines bright sub-components even at relatively large separation,
whilst keeping faint sources as single objects. 
As is shown in Figure~\ref{fig:grouphist} (dotted line), this
procedure does not seriously affect the number of doubles and triples,
while strongly decreasing the number of spurious
aggregations. 

\begin{figure}
\vspace{8cm}  
\includegraphics{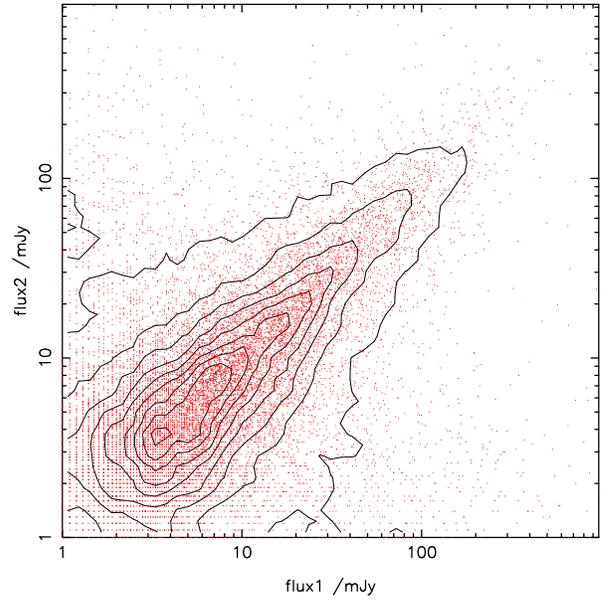} 
\caption{Distribution of flux densities for members of `double sources'
  delineated by the combining procedure described in the text.
\label{fig:f1f2} }
\end{figure}

As a further step in identifying groups of sources that should be
combined, for each `double source' identified, we plotted 
the flux of the components against each 
other (Figure~\ref{fig:f1f2}).
The two fluxes are highly correlated, with most pairs lying within a
narrow band about the value $flux1/flux2=1$.
Such a correlation is expected for real double radio sources, for
which the flux from the two lobes is correlated. 
Source pairs which our linking criterion does not combine show no
correlation between the fluxes of the two components. 
This flux correlation suggests a further criterion to restrict the
combination of sources to only physically associated objects: we
combine pairs of sources only if their fluxes differ by a factor less
than 4 (i.e. $1/4\le\frac{flux1}{flux2}\le4$). Although somehow ad hoc,
this upper limit on the flux ratio seems to be a reasonable guess
given that it delimits a region on the $flux1-flux2$ plane that, even
though quite narrow (as it has to be for physically associated
objects), encloses roughly 95\% of the double sources as found
through equation \ref{eqn:rlink}. Stability of clustering
measurements as a function of the collapsing procedure will be
explored elsewhere.

The final number of groups obtained by including this further
constraint is 17554 and their distribution as a function of number of
members is shown by the dashed line in Figure~\ref{fig:grouphist}.
While keeping almost all the doubles and triples, the number of
spurious groups formed by multicomponents is now reduced to a very
small fraction ($<10^{-2}$) of the whole catalogue.
In this way we ended up with a catalogue C3 formed by 167,433 sources.

\subsection{Incompleteness}
Another important issue, noted by White et al. (1997), is
the incompleteness of the survey. Becker, White \& Helfand (1995)
estimate the catalogue to be $95\%$ complete at 2 mJy and $80\%$ 
complete at 1 mJy.
This could affect our results in two ways. First, if there are any
spatial inhomogeneities in completeness, spurious clustering will be
introduced. Second, even if the incompleteness is uniform, the
redshift distribution will be different from a complete
flux-limited sample, and different 
in an unpredictable way. This difference will change the relation between 
2-d and 3-d
quantities, and so bias the 3-d measurements, as discussed in Section
4. 

\begin{figure}
\vspace{8cm}  
\includegraphics{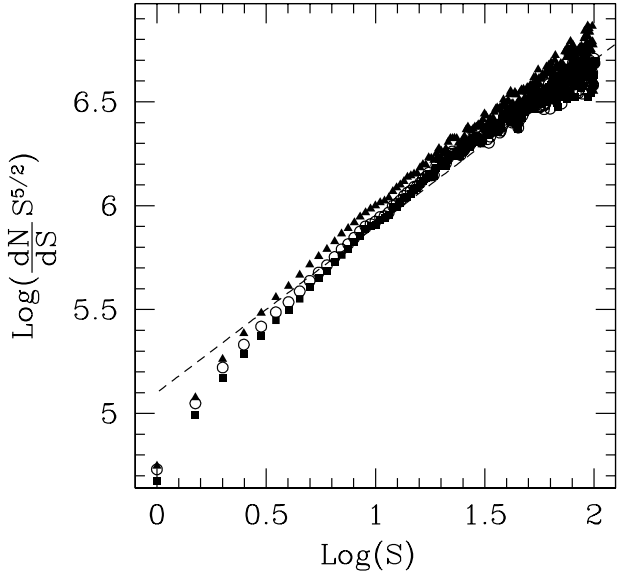} 
\caption{Differential number counts versus the integrated flux density
  for the three catalogues C1 (filled triangles), C2 (filled squares) and
  C3 (open circles), obtained from the original FIRST survey as
  explained in Sections 2.1 and 2.2. The dashed line shows the
  $\frac{dN}{dS}$ distribution as determined by Windhorst et
al. (1985). 
\label{fig:ncount} }
\end{figure}

In Figure~\ref{fig:ncount} we plot the differential source counts
normalised to the distribution expected for a non-evolving population
in Euclidean space vs the integrated flux density for the three
catalogues C1 (filled triangles), C2 (filled squares) and 
C3 (open circles), obtained from the original FIRST survey as
described in Sections 2.1 and 2.2. The dashed line shows the
power-law fit to the $\frac{dN}{dS}$ distribution determined by
Windhorst et al. (1985). 
There is a drop in the number counts for
fluxes smaller that 3~mJy, and this lack of faint objects becomes
very important below 2~mJy. Furthermore the summation of fluxes
in the recombination of sub-components affects the 
incompleteness of the resulting catalogues. 
This is especially important in the faint-flux region. 
Note that as we chose an initial threshold of 1 mJy, we may lose
pairs with $f_1+f_2>3$ mJy, but with either $f_1$ or $f_2$ $<1$ mJy.

 From Figure~\ref{fig:ncount}, it is a reasonable assumption that the 
survey is complete at fluxes greater than 3~mJy. 
Hence we set a flux limit of 3~mJy for the catalogues used in our
analyses.
At this flux limit the C1,C2 and C3 samples contain respectively
101787, 83287 and 86074 objects.
In the rest of the paper we adopt the C3 version of the catalogue, but 
we compare to the results with
those obtained for C1 and C2.

\section{CLUSTERING ANALYSIS}

\subsection {The Angular Two-Point Correlation Function} 
Correlation-function analysis has become the standard way to quantify
the clustering of different populations of astronomical sources.
Specifically the angular two-point correlation function
$w_{12}=w(\theta)$ gives the excess probability, with respect to a
random Poisson distribution, of finding two sources in the solid
angles $\delta\Omega_1$ $\delta\Omega_2$ separated by an angle
$\theta$, and it is defined as
\begin{eqnarray} 
\delta P=n^2\delta\Omega_1\delta\Omega_2\left[1+w(\theta)\right]
\label{eqn:wthetadef} 
\end{eqnarray}
where $n$ is the mean number density of objects in the catalogue under
consideration.
\begin{figure}
\vspace{8cm}  
\includegraphics{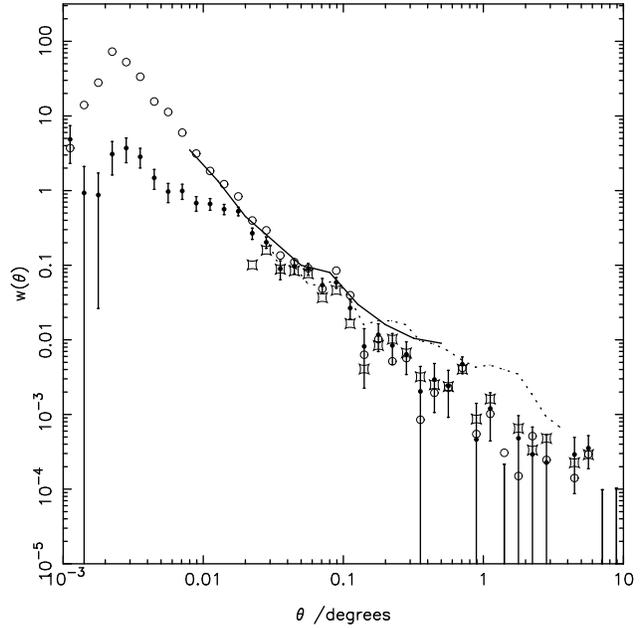} 
\caption{The angular correlation functions $w_h$ for C1 (open
circles), C2 (open squares) and C3 (filled circles). The error bars show
Poisson estimates for the C3 points. The lines show $w$ measurements from
Cress et al. (1996): the solid line is for the raw catalogue and the
dotted line for their `collapsed' (component-combined) catalogue.
\label{fig:wplot} }
\end{figure}

We measured $w$ for the three catalogues C1, C2 and C3 using various
techniques. First we generated a random catalogue of positions and
fluxes and selected sources above the sensitivity limit from the 
coverage map. Then we counted the number of distinct data-data pairs ($DD$),
data-random pairs ($DR)$, and random-random pairs ($RR$) as a function
of angular separation. 
We then measured  $w$ from the three commonly-used estimators (Hamilton, 
1993; Landy \& Szalay, 1993; Peebles, 1980), \begin{eqnarray} 
w_{h} = &\frac{4DD \ RR}{DR^2}  &-1  \nonumber \\
    w_{ls} = &\frac{DD + RR -DR }{DR} &-1 \nonumber\\
   w_{dr} = &\frac{2 DD}{DR} &-1   
\label{eqn:wtheta_ests} 
\end{eqnarray}
We also counted the number of sources in equal-area cells as discussed
in section 3.2 and estimated $w$ from the cell counts, 
\begin{eqnarray} 
 w_{cell}  = \frac{ \langle N_1 N_2 \rangle }{\langle N_1 \rangle\langle N_2
\rangle} - 1 
\label{eqn:wthetc_cell} 
\end{eqnarray}

For all catalogues, $w_{ls}$ and $w_{h}$ were essentially identical, and
not significantly different from $w_{cell}$. The estimate $w_{dr}$ agrees
perfectly on scales $\theta \simlt 1^\circ$, but 
is slightly lower at $\theta \simgt 1^\circ$. 

In Figure~\ref{fig:wplot} we show $w_h$ for the three catalogues C1,
C2 and C3; the error bars show Poisson estimates for the C3
points. As the distribution is clustered these estimates
only provide a lower limit to the real errors (see e.g. Mo et al.,
1992) that might be obtained for instance by using the bootstrap
re-sampling method. More precise estimates will be given in section 3.2
when we will move on to more quantitative results.\\
Note that, as it will be extensively discussed later in the paper,
versions of the same catalogue obtained by adopting different
collapsing procedures lead to different estimates of the angular
correlation function; as it can be seen by comparing the C2
measurement with those given by the analysis of the C1 and C3
catalogues, it turns out that the higher is the number of objects that
have been combined together, the shallower is the slope of the
corresponding $w_h$.
In more detail the C1 measurement shows a steepening at $\theta \simlt
0.02^\circ$ caused by the over-counting of multiple-component
sources. At very small angles ($\theta \simlt 0.002^\circ$) 
the amplitude drops to $-1$, as close pairs of sources become unresolved
by the survey. 
The C2 measurement is $-1$ for $\theta < 0.02^\circ $ because all $DD$
pairs have been removed from the catalogue by the recombining procedure.
Our C2 measurement should be comparable to the $w$ estimate from 
Cress et al. (1996). In fact our measurements for both C1 and C2 are in
very good agreement with Cress et al. (1996)
for $\theta < 0.1^\circ$, but at larger scales we find a slightly
steeper slope, leading to a difference of a factor of 2 by 
$\theta \sim 2^\circ$. 
The discrepancy is marginally significant given the estimated errors,
and is probably due to using the more recent version of the catalogue. 

The C3 measurement continues approximately as a power law to the limit
of the survey resolution, $\theta \sim 0.002^\circ$. This gives us some
confidence that our multi-component recombining procedure is
reliable.

\subsection{Variance} 
The galaxy distribution function ({\it counts in cells}) gives the
probability of finding $N$ galaxies in a cell of particular size and
shape. In principle the counts-in-cells distribution  is
straightforward to compute; one divides the space into cells of a given
volume (or in our case, of given area) and shape, and then counts the
number of galaxies in each cell. Obviously there cannot be less than
zero galaxies in a cell, but the number of objects in each cell can
grow with no upper bound (unless some negative feedback process takes
place).

If we define the $k$th moment of the counts as
\begin{eqnarray}
\mu_k=\left<(N-\bar{N})^k\right>
\label{eqn:mu_k}
\end{eqnarray}
where $\bar{N}=n \Omega$ is the mean count in the solid angle
$\Omega$, it is then possible to write these moments in terms of the
$n$-point correlation functions (see e.g. Peebles, 1980). 
The second moment of the galaxy distribution function
is related to the two-point correlation function $w_{12}$ through the
expression
\begin{eqnarray}
\mu_2=\bar{N}+(\bar{N})^2\Psi_2
\label{eqn:mu_2}
\end{eqnarray}
where $\bar{N}$ is the shot noise resulting from the discrete nature
of the sources (Poisson noise), and
\begin{eqnarray}
\Psi_2\equiv\frac{1}{\Omega^2}\int w_{12}\:d\Omega_1\:d\Omega_2
\label{eqn:Psi_2}
\end{eqnarray}
is the normalised variance in terms of the two-point correlation
function integrated over a cell of area $\Omega$ 
and particular shape.

By assuming a power-law form 
\begin{eqnarray}
w(\theta)=A\;\theta^{1-\gamma},
\label{eqn:w_power}
\end{eqnarray}
and by considering square cells of size $\Omega=\Theta\times\Theta$
square degrees, we can evaluate the integral in equation \ref{eqn:Psi_2} (see
Totsuji \& Kihara, 1969), obtaining
\begin{eqnarray}
\sigma^{2}\equiv\frac{\mu_2-\bar{N}}{(\bar{N}/\Omega)^2}=\int
A\:\theta^{1-\gamma}d\Omega_1d\Omega_2=
A\:C_{\gamma}\Theta^{5-\gamma}
\label{eqn:sigma2}
\end{eqnarray}
where $C_{\gamma}$ is a coefficient depending on $\gamma$ which can be
evaluated numerically by Monte Carlo methods (see e.g. Lahav \& Saslaw,
1992). It is therefore possible to use the 
$\sigma^2 - \Theta$ relation to evaluate the
two parameters ($A$, $\gamma$) which describe the correlation function.

\begin{figure}
\vspace{8cm}  
\includegraphics{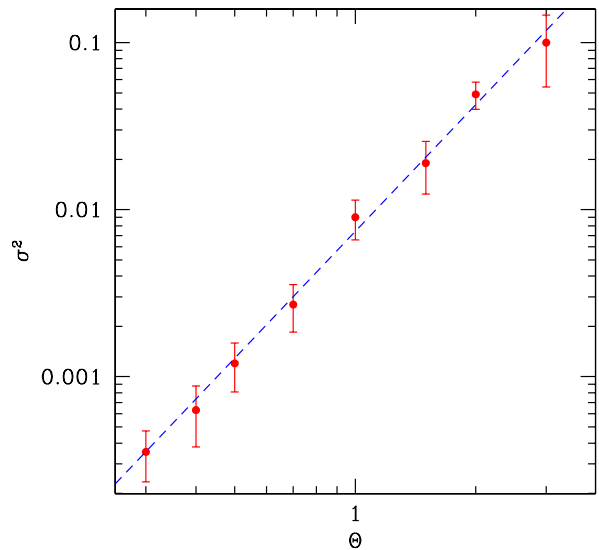} 
\caption{The normalised variance $\sigma^{2}$ vs the cell size $\Theta$
  for our C3 catalogue of the FIRST survey. Errors are 
estimated from the variance in four random subsets.
\label{fig:sig_theta3} }
\end{figure}

To carry out our analysis we first project the coordinates of
the objects in the survey onto an {\it equal-area projection}.  
We then divide the region into square
cells of constant $x$ and $y$ in rectangular coordinates for the
projection, and count the number of objects in each cell.
We use the coverage map to discard all cells which are
partially occupied, as described in section 2.2.
The procedure is  repeated for different cell sizes, from
$\Delta x=\Delta y=\Theta=0.3^\circ$ up to $3^\circ$; at each cell size, 
the variance $\sigma^2$ is calculated from the estimator in 
equation~\ref{eqn:sigma2}.

Figure~\ref{fig:sig_theta3} shows
$\sigma^{2}$ as a function of $\Theta$ for the C3 catalogue.
The slope and the intercept of the plot are estimated by
a least-squares procedure minimising the quantity
\begin{eqnarray}
 \chi^2(a,b)=\sum_{i=1}^{N}\left(\frac{\log(\sigma)_i-a-b\log(\Theta)_i}{\Delta_i}\right)^2
\label{eqn:chi2}
\end{eqnarray}
with $b\equiv(5-\gamma)$, $a\equiv
\log(AC_{\gamma})$. The errors $\Delta_i$ are
obtained using the `partition bootstrap method' in which the
normalised variance is calculated for four subdivisions of the survey
region and the standard deviation of these measurements at each angle
is used as a measure of the error. 
Note that this is not strictly a $\chi^2$ statistic, since the
individual points are not independent. From this analysis we 
find
\begin{eqnarray}
\gamma=2.50\pm 0.1 \;\;\; ; \;\;\; A=(1.06\pm 0.1)\cdot 10^{-3},
\label{eqn:sig2_fit3}
\end {eqnarray}
where $C_{\gamma}=6.52$ has been obtained by solving the integral in
equation~\ref{eqn:sigma2} with the
condition $\theta\ge 0.01^\circ$, appropriate to the component-combining 
algorithm used to construct C3.

The amplitude $A$ is
smaller by about one order of magnitude than the values
obtained from optical surveys (e.g. $A=(3.82\pm
0.12)\cdot 10^{-2}$ for the APM survey (Maddox et al., 1990)); 
this is due to the `washing out' of
structure by the wide redshift span of radio surveys. 
The value we find for $\gamma$, high in 
comparison with optical surveys
(e.g. $\gamma\sim 1.8$, Peebles (1980), Maddox et al. (1990)), is 
discussed in Section 4.

\begin{figure}
\vspace{8cm}  
\includegraphics{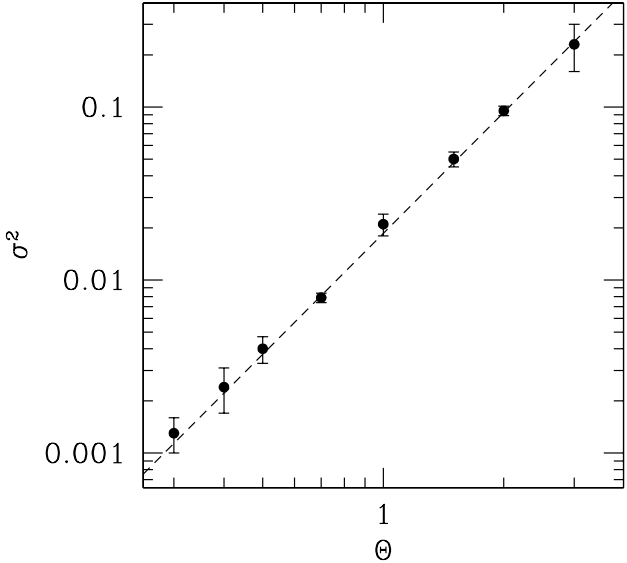} 
\caption{The normalised variance $\sigma^{2}$ vs cell size $\Theta$
  for the C1 catalogue of the FIRST survey. Errors are 
estimated from
  the variance in four random subsets.
\label{fig:sig_theta1} }
\end{figure}
\begin{figure}
\vspace{8cm}  
\includegraphics{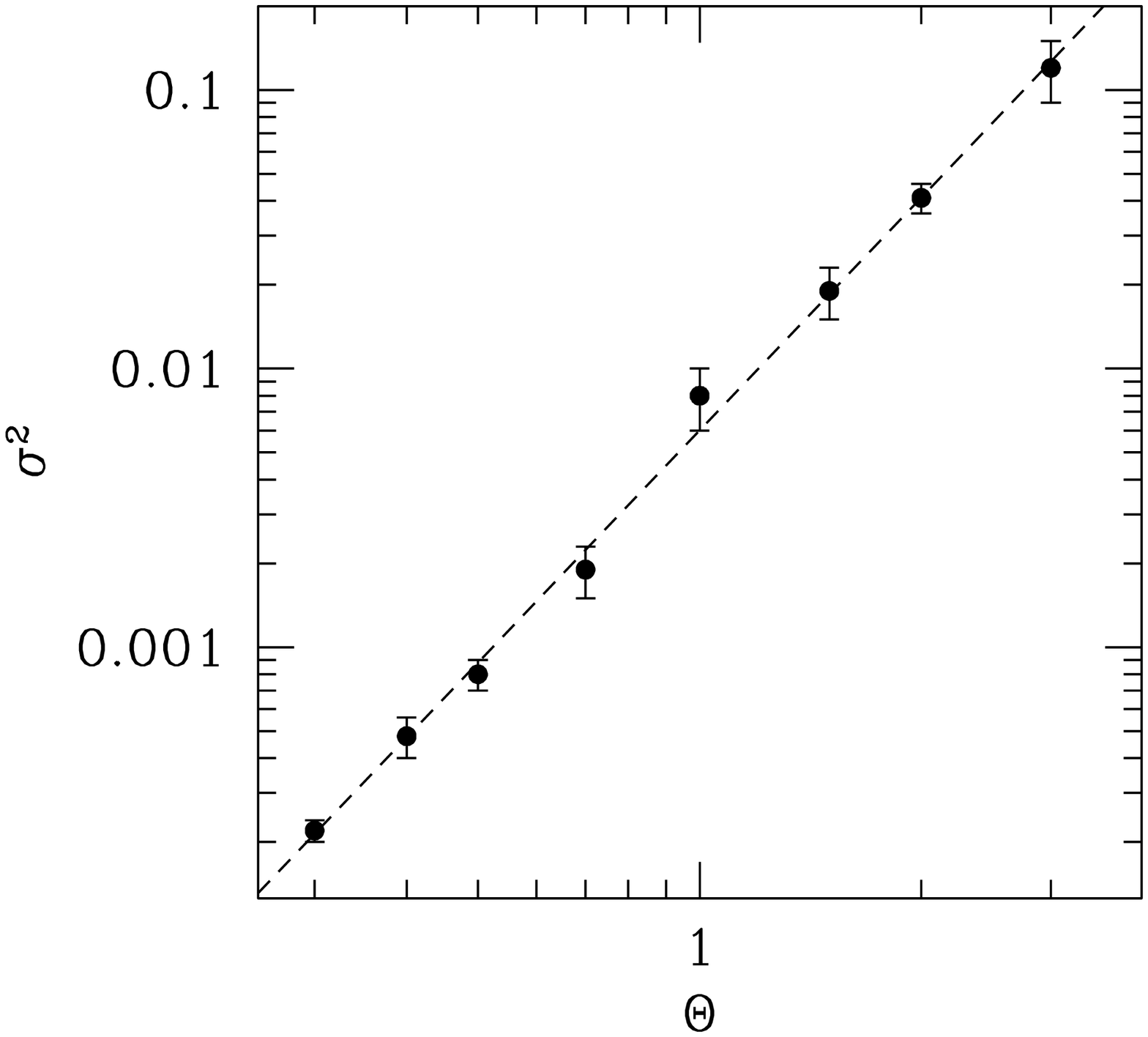} 
\caption{The normalised variance $\sigma^{2}$ vs cell size $\Theta$
    for the C2 catalogue of the FIRST survey. Errors are estimated from
  the variance in four random subsets.
\label{fig:sig_theta2} } 
\end{figure}

By repeating the analysis for the C1 and C2 samples 
(Figures~\ref{fig:sig_theta1} and \ref{fig:sig_theta2}), 
we find respectively:
\begin{eqnarray}
\gamma=2.68\pm 0.07 \;\;\; ; \;\;\; A=(1.52\pm 0.06)\cdot 10^{-3}
\label{eqn:sig2_fit1}
\end {eqnarray}
and
\begin{eqnarray}
\gamma=2.23\pm 0.1 \;\;\; ; \;\;\; A=(1.42\pm 0.1)\cdot 10^{-3}.
\label{eqn:sig2_fit2}
\end {eqnarray}
To work out the latter value of $A$, $C_{\gamma}$ has been
calculated from equation~\ref{eqn:sigma2} 
by imposing the condition $\theta\ge 0.02^\circ$, again from
the nature of the component-combining criterion. 
The values in equation~\ref{eqn:sig2_fit2} are in close
agreement with those obtained by Cress et al. (1996).\\

Comparison of the slope $\gamma$ in equations~\ref{eqn:sig2_fit3},
\ref{eqn:sig2_fit1}  and \ref{eqn:sig2_fit2} shows
how its value decreases according to the number of source-components 
combined in the different catalogues ($\gamma$ largest for no
combining). The 
presence of more `sources' or `source components' in close proximity, 
regardless of their 
physical association, results in a stronger correlation on smaller scales.

\subsection{Skewness}
Further information on clustering is contained in the 
higher moments of the
distribution, such as the skewness and the kurtosis. Assuming 
Gaussian primordial perturbations and linear theory, both the skewness 
and the kurtosis remain zero (Peebles, 1980). A detection of a non-zero 
value for these quantities may then be evidence of non-linear
gravitational clustering.
Departures from Gaussian distributions might also reflect
non-Gaussian initial conditions predicted in some cosmogonies such as
cosmic-string and texture models. Detection of non-zero values for 
skewness and kurtosis is therefore
of fundamental importance.

\begin{figure}
\vspace{8cm}  
\includegraphics{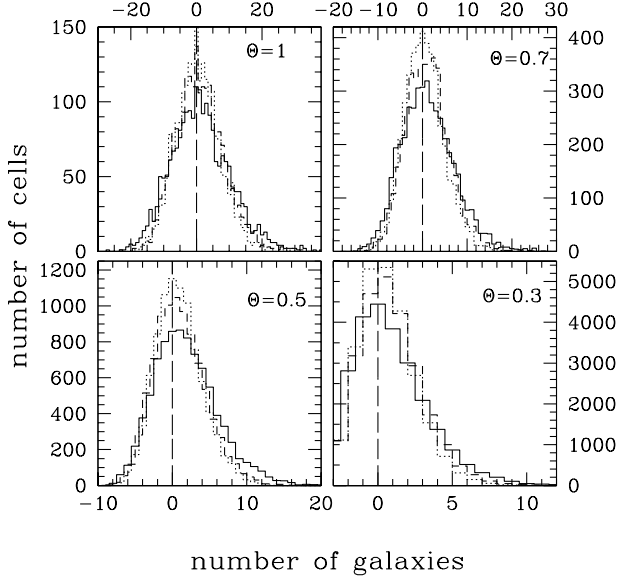} 
\caption{Distribution function for the FIRST survey for
  different cell sizes. The solid lines show the results obtained
  for C1, and the dotted and dashed lines are
  for C2 and C3 respectively. The $x$ axes have been rescaled according 
to $x'=x-x_m$, where $x_m$ is the median. 
\label{fig:nhist} }
\end{figure}

\begin{figure}
\vspace{8cm}  
\includegraphics{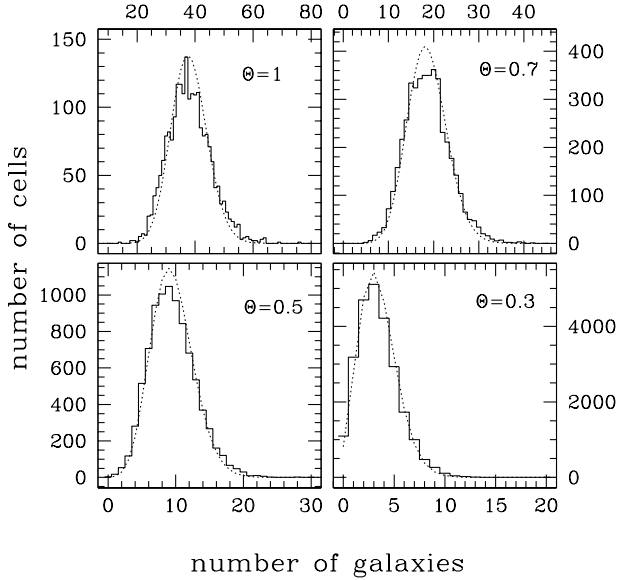} 
\caption{Distribution function for the FIRST survey for
  different cell sizes. The solid line shows the results for sample 
C3 compared to a Poisson distribution (dotted line). 
\label{fig:ranhist} }
\end{figure}

Kurtosis is not considered here
because its estimated value is dominated by noise.
Skewness is actually
observed in each of the catalogues C1, C2 and C3 under consideration, 
starting from angles $\Theta\sim 1^\circ$. As seen in Figure~\ref{fig:nhist},
the galaxy distribution functions are noticeably skewed towards larger
values of source-numbers, the tail becoming increasingly
prominent as the cell-size becomes
smaller. To illustrate, Figure~\ref{fig:ranhist} shows the comparison between
sample C3 and a Poisson distribution.
K-S tests show that the probabilities that the histograms are
consistent with Poisson distributions are $ < 0.01$\%.


Using an approach similar to that described in Section 3.2,
the third moment (skewness) is related to the
three-point correlation function $\omega_{123}$ through the expression
\begin{eqnarray}
\mu_3=\bar{N}+3(\bar{N})^2\Psi_2+(\bar{N})^3\Psi_3,
\label{eqn:mu_3} 
\end{eqnarray}
where
\begin{eqnarray}
\Psi_3\equiv\frac{1}{\Omega^3}\int\omega_{123}\:
d\Omega_1\:d\Omega_2\:d\Omega_3
\label{eqn:Psi_3} 
\end{eqnarray}
is the normalised skewness in terms of the three-point correlation
function integrated as in equation~\ref{eqn:Psi_2}.

An estimator for the normalised variance is
\begin{eqnarray}
\Psi_2\equiv\frac{\mu_2-{\bar N}}{({\bar N})^2}=\sigma^2\Omega^2 \;,
\label{eqn:Psi_2_again} 
\end{eqnarray}
and an estimator for the normalised skewness is
\begin{eqnarray}
\Psi_3\equiv \frac{\mu_3-{\bar N}-3(\bar{N})^2\Psi_2}{(\bar{N})^3} \;.
\label{eqn:Psi_3_again} 
\end{eqnarray}
It was shown
(Peebles, 1980; Juszkiewicz et al., 1995; Coles \& Frenk, 1991)
that if the initial perturbations were Gaussian 
and they grew with cosmic time purely due to gravity, 
then in second-order perturbation theory 
(i.e. in the quasi-linear regime, 
$\delta \equiv {\delta \rho \over \rho } \sim 1$)  
\begin{eqnarray}
S_3^*(R) \equiv 
{ \langle \delta^3 \rangle_R \over \langle \delta^2 \rangle^2_R } 
=  { 34 \over 7} - (n+3)\;, 
\label{eqn:s_3star} 
\end{eqnarray}
where $S_3^*(R)$ is 
the spatial normalized 
skewness in randomly placed spheres of radius $R$
and $n$ is the index of the primordial
power-spectrum of fluctuations.  
It follows that in this case the projected quantities obey
$\Psi_3\propto (\Psi_2)^{\alpha}$, with $\alpha=2$.
For local surveys (e.g. Peebles, 1980), $\alpha=2$ is also expected 
from the empirical relation between the 
3-point and 2-point correlation functions:
\begin{eqnarray}
\zeta({\bf r}_1, {\bf r}_2)=Q [ \xi(r_1)\xi(r_2)+\xi(r_1)\xi(r_{12})+
\nonumber \\+\xi(r_2)\xi(r_{12}) ] \propto \xi^2(r),
\label{eqn:zeta}
\end{eqnarray}
where $\zeta({\bf r}_1, {\bf r}_2)$ is the spatial three-point 
correlation function.

\begin{figure}
\vspace{8cm}  
\includegraphics{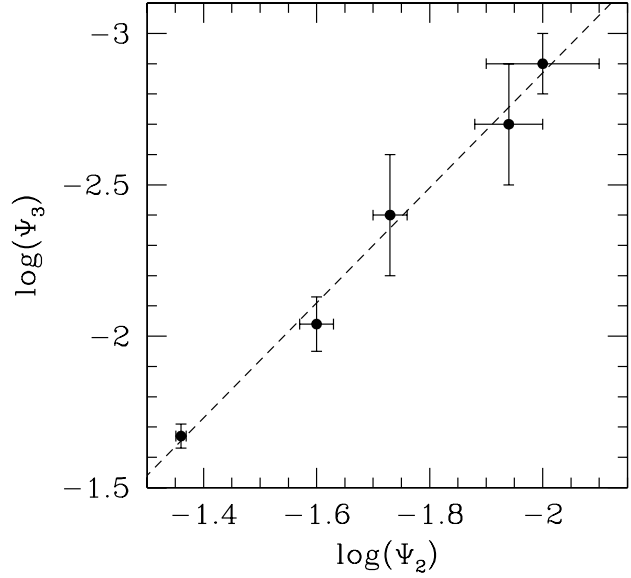} 
\caption{The normalised variance $\Psi_2=\Omega^2\sigma^2$ vs the
  normalised, dimensionless skewness $\Psi_3$ for C3. Errors have 
been calculated as in Section 3.2.
\label{fig:psi2psi3_3}  }
\end{figure}
\begin{figure}
\vspace{8cm}  
\includegraphics{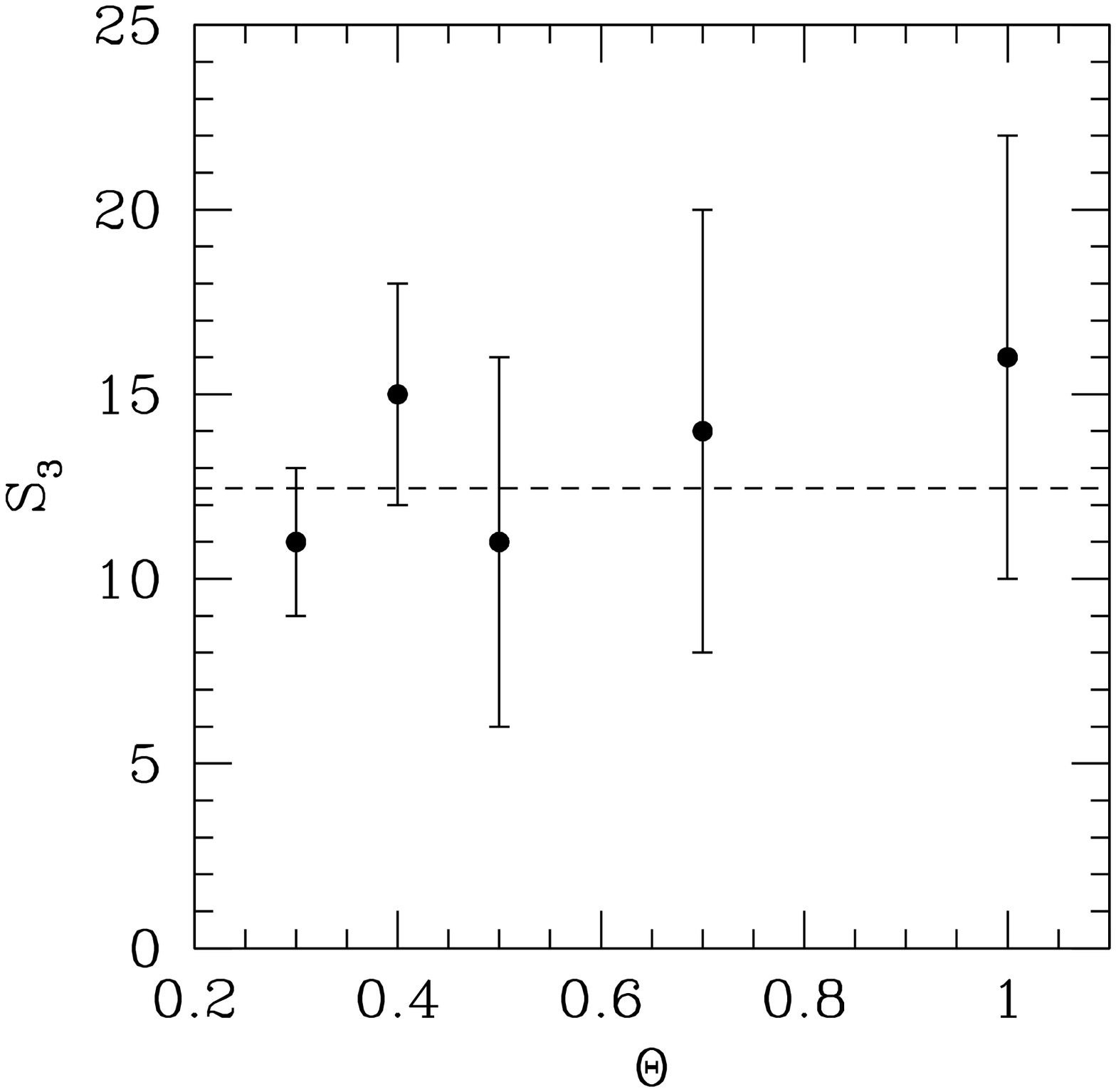} 
\caption{$S_3$ as a function of $\Theta$ for C3, 
assuming $\alpha=2$. Errors are
  estimated as in Section 3.2.
\label{fig:S3_3} }
\end{figure}
\begin{figure}
\vspace{8cm}  
\includegraphics{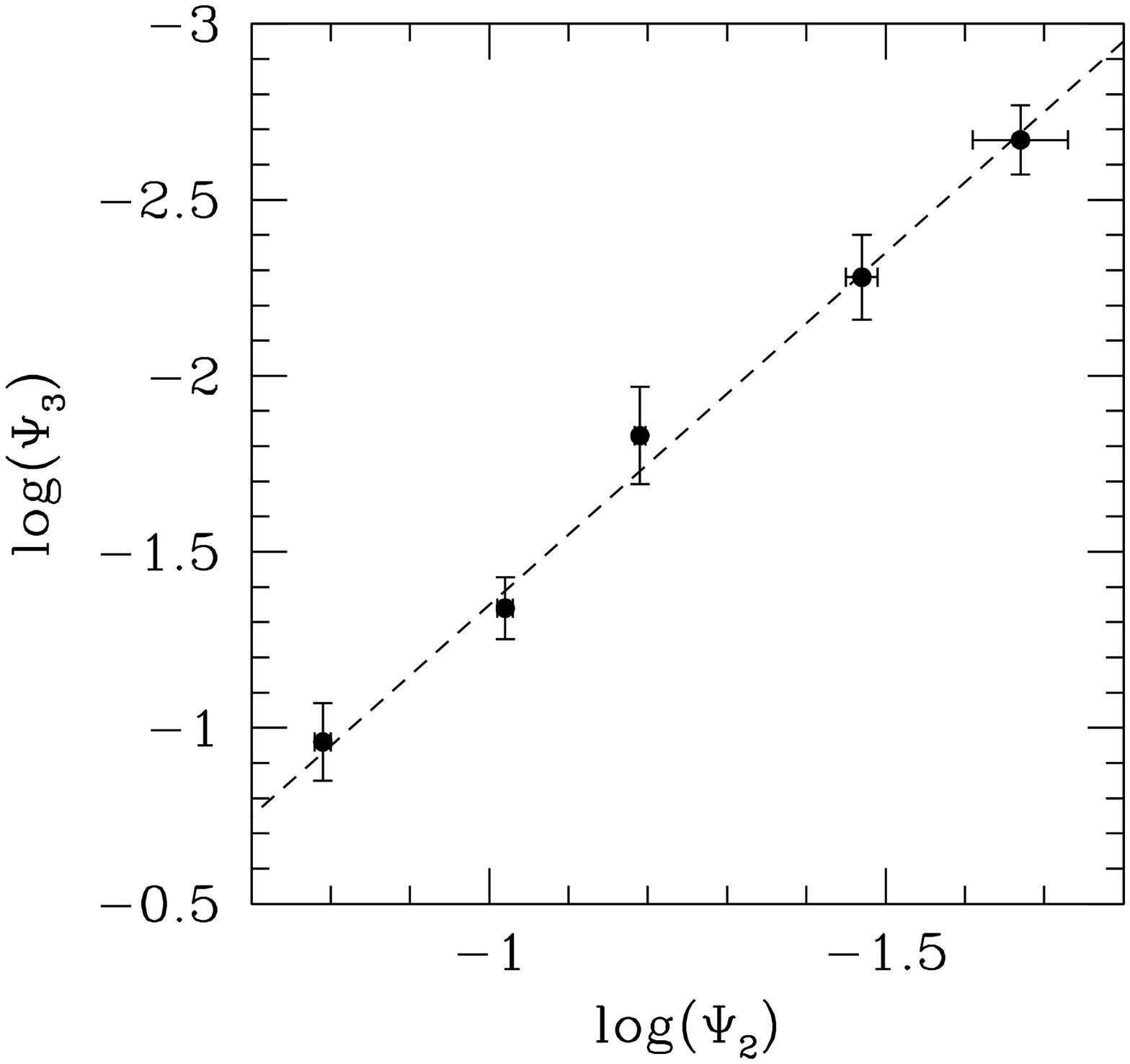} 
\caption{The normalised variance $\Psi_2=\Omega^2\sigma^2$ vs the
  normalised, dimensionless skewness $\Psi_3$ for C1. Errors have 
been calculated as in section (3.2).
\label{fig:psi2psi3_1} }
\end{figure}
\begin{figure}
\vspace{8cm}  
\includegraphics{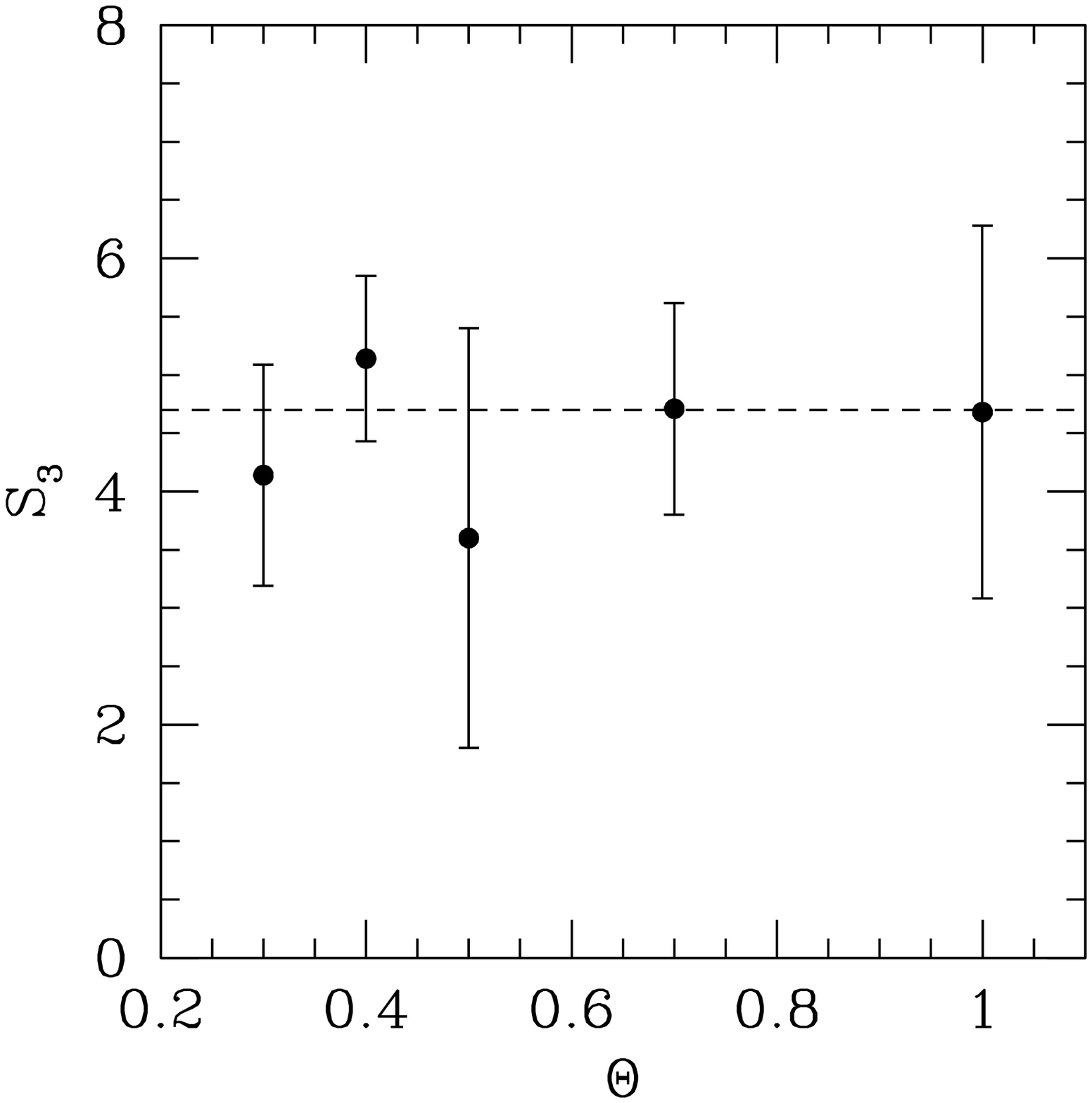} 
\caption{$S_3$ as a function of $\Theta$ 
for C1 assuming $\alpha=2$. Errors are
  estimated as in Section 3.2.
\label{fig:S3_1} }
\end{figure}
\begin{figure}
\vspace{8cm}  
\includegraphics{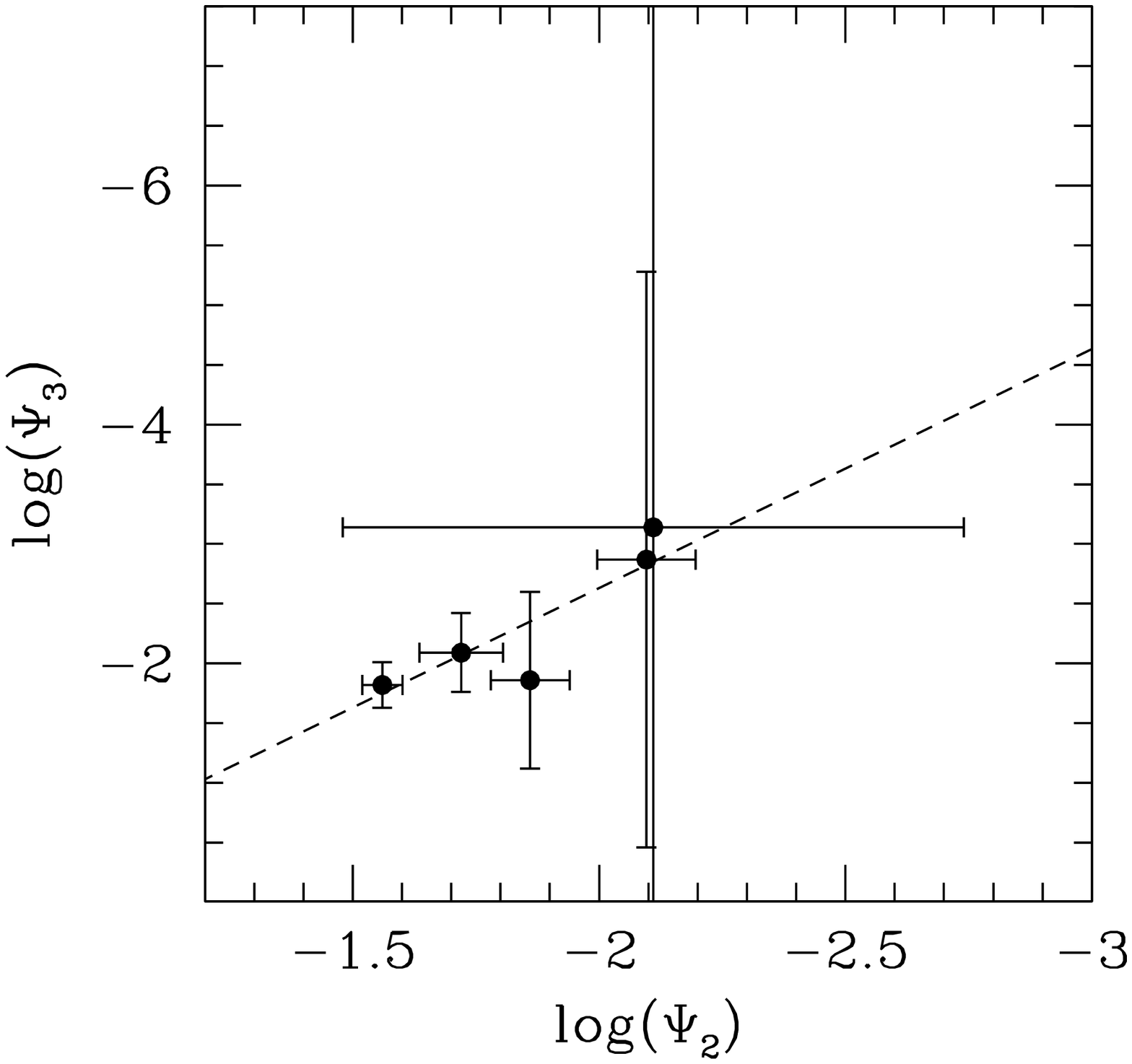} 
\caption{The normalised variance $\Psi_2=\Omega^2\sigma^2$ vs the
  normalised, dimensionless skewness $\Psi_3$ for C2. Errors have 
been calculated as in Section 3.2.
\label{fig:psi2psi3_2} }
\end{figure}
\begin{figure}
\vspace{8cm}  
\includegraphics{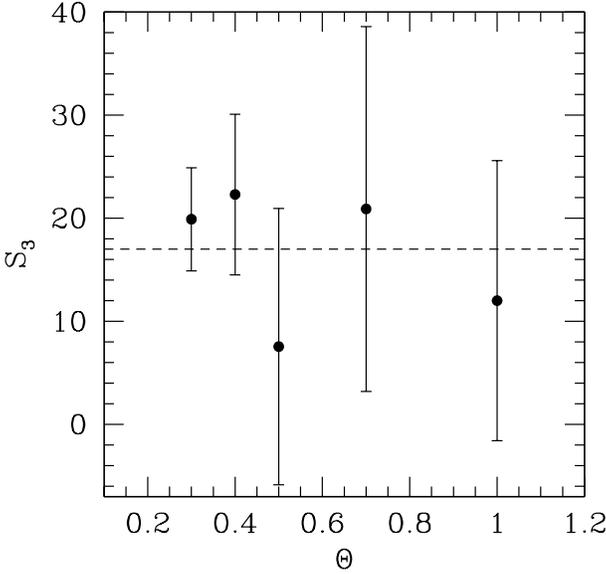} 
\caption{$S_3$ as a function of $\Theta$ for C2 assuming 
$\alpha=2$. Errors are
  estimated as in Section 3.2.
\label{fig:S3_2} }
\end{figure}

Figure~\ref{fig:psi2psi3_3} shows $\log(\Psi_3)$
vs $\log(\Psi_2)$ for angles $0.3^\circ\le\Theta\le1^\circ$; a 
least-squares fit relates the quantities by the form
\begin{eqnarray}
\Psi_3=S_3(\Theta) \; \Psi_2^{\alpha}
\label{eqn:Psi3_Psi2} 
\end{eqnarray}
with $\alpha=1.9\pm 0.1$ and $S_3(\Theta)=9\pm 3$ and errors 
estimated as in Section 3.2. Therefore 
the value found for $\alpha$
supports the assumption of 
gravitational growth of perturbations 
from Gaussian initial conditions.  

To test the constancy of $S_3$ as predicted by the hierarchical
theory, we plot this quantity as a function of the cell
size $\Theta$ on the assumption of $\alpha=2$ (Figure~\ref{fig:S3_3}).
The plot supports the hypothesis of constant $S_3(\Theta) = 12.5 \pm 1.5$.

The same analysis for the catalogue C1 
(containing all sources), and C2 (generated by combining `sources' only 
according to their separation), finds: 
$\alpha=2.0\pm0.1,\;S_3(\Theta)=4\pm 2$;
$S_3(\Theta)=4.7\pm 0.3$ for C1 (see
Figures~\ref{fig:psi2psi3_1} and \ref{fig:S3_1}), 
 and $\alpha=1\pm2,\;S_3(\Theta)=1\pm 3$;
$S_3(\Theta)=17\pm 5$ for C2.  In this latter
case the estimate of the quantities
is dominated by noise (see Figures~\ref{fig:psi2psi3_2}
and \ref{fig:S3_2}). 
\begin{table*}
\caption{COUNTS IN CELLS ANALYSIS RESULTS}
\begin{center}
\begin{tabular}{|l|l|l|l|} 
\hspace{0.5cm}&{\bf{C1}}&{\bf{C2}}&{\bf{C3}}\\
\hline
$\gamma$&$2.68\pm 0.07$&$2.2\pm 0.1$&$2.50\pm 0.1$\\
\hline
$A$&$(1.52\pm 0.06)\cdot 10^{-3}$&$(1.4\pm
0.1)\cdot10^{-3}$&$(1.06\pm 0.1)\cdot 10^{-3}$\\
\hline
$\alpha$&$ 2.0\pm 0.1$&$1\pm 2$&$1.9\pm 0.1$\\
\hline
$S_3(\Theta)$&$4\pm 2$&$1\pm 3$&$9\pm3$\\
\hline
$S_3(\Theta)_{\alpha=2}$&$ 4.7\pm 0.3$&$17\pm 5$&$12.5\pm 1.5$\\
\hline
\end{tabular}
\end{center}
\end{table*}
The results are highly sensitive to the process of `combining' 
multi-component sources. Combining two or more `sources' in close 
proximity strongly affects the distribution function in two ways. It 
removes the tail, thereby reducing skewness, and at the same time it 
reduces the variance as more cells will contain numbers close to the 
average of the distribution. The effects become rapidly more important as 
the number of combined sources increases; this explains the remarkable 
difference in the values obtained from the three catalogues. In 
particular the results for the C2 catalogue show that the variance and the 
skewness of its distribution become so small as to be dominated by noise. 
Note that the differences between C2 and C3 are due to less than 3000 
`objects' ($\sim 4\%$). 

Table 1 summarises the results obtained from the counts-in-cells
analysis for the three catalogues C1, C2 and C3.

\section{RELATION TO SPATIAL QUANTITIES}

\subsection{The Spatial Correlation Function}
Once both the 
cosmological model and the selection function of the sample are 
specified, the standard way of relating the spatial ($3d$) correlation 
function $\xi(r,z)$ to the angular ($2d$) one $w(\theta)$
is via the relativistic Limber equation (Peebles, 1980). For 
Einstein-de Sitter universe ($\Omega=1, \Lambda=0$),
\begin{eqnarray}
w(\theta)=2\:\frac{\int_0^{\infty}\int_0^{\infty}x^4\Phi^2(x)\xi(r,z)dx\:du}{\left[\int_0^{\infty}x^2\Phi(x)dx\right]^2},
\label{eqn:limber} 
\end {eqnarray}
where $x$ is the comoving coordinate
\begin{eqnarray}
x=\frac{c}{2 H_0}\left[1-(1+z)^{-1/2}\right], 
\label{eqn:xcomoving} 
\end{eqnarray}
and the selection function $\Phi(x)$
satisfies the relation
\begin{eqnarray}
{\cal N}=\int_0^{\infty}\Phi(x) x^2 dx=\frac{1}{\Omega_s}\int_0^{\infty}N(z)dz,
\label{eqn:Ndense} 
\end{eqnarray}
in which $\cal N$ is the mean surface density on a surface of solid angle
$\Omega_s$ and $N(z)$ is the number of objects in the given survey
within the shell ($z,z+dz$).

If we assume for $\xi(r,z)$ a power-law redshift-dependent form
\begin{eqnarray}
\xi(r,z)=\left(\frac{r}{r_0}\right)^{-\gamma}(1+z)^{-(3+\epsilon)},
\label{eqn:xi_power} 
\end{eqnarray}
$\gamma$ constant with $z$, where $r$ is the proper coordinate, $r_0$ the 
correlation scale
length at redshift $z=0$ and $\epsilon$ a parameter describing the
redshift evolution of the spatial correlation function, then, in
comoving coordinates $r_c=r(1+z)$, $\xi$ assumes the form
\begin{eqnarray}
\xi(r,z)=\left(\frac{r_c}{r_0}\right)^{-\gamma}(1+z)^{\gamma-(3+\epsilon)}.
\label{eqn:xi_power_com} 
\end{eqnarray}
Specific values for $\epsilon$ have the following significance:
$\epsilon=0$ implies constant clustering in proper coordinates;
$\epsilon=\gamma-3$ implies constant clustering in comoving
coordinates; and $\epsilon=\gamma-1$ represents growth of clustering under
linear theory (Peebles, 1980; Treyer \& Lahav, 1996).

In the small-angle approximation
$r\simeq(u^2+x^2\theta^2)^{1/2}/(1+z)$ we then find that  the amplitude A defined in equations
\ref{eqn:w_power} and \ref{eqn:sigma2} is given by the expression (e.g. Loan et al., 1997; Peebles, 1980):
\begin{eqnarray}
\label{eqn:A_power} 
A =\left(\frac{r_0H_0}{c}\right)^{\gamma}2^{1-\gamma}\frac{\Gamma\left(\frac{1}{2}\right)\Gamma\left(\frac{\gamma-1}{2}\right)}{\Gamma\left(\frac{\gamma}{2}\right)}\times\\
  \nonumber
\times
  \frac{\int_0^{\infty}\left[1-(1+z)^{-1/2}\right]^{1-\gamma}N^2(z)(1+z)^{\gamma-\frac{3}{2}-\epsilon}dz}{\left[\int_0^{\infty}N(z)dz\right]^2}
\end{eqnarray}
where $\Gamma$ is the Gamma function. This expression, considering
the functional form (\ref{eqn:w_power}) for
 $w(\theta)$,  directly fixes the value of $r_0$ once the evolution
parameter $\epsilon$ is given.

\subsection{The Spatial Skewness}
An analysis analogous to that in Section 4.1 provides
the 3-dimensional value for the skewness 
$S_3^*(R)$ in randomly-placed spheres of comoving  radius $R$. 
This is related to
the projected (2-dimensional) quantity $S_3(\Theta)$, 
in the case of circular cells of radius $\theta$, through the equation
(Gazta\~naga, 1995)
\begin{eqnarray}
S_3(\Theta)=r_3 S_3^*(R)\left(\frac{C_3}{B_3}\right)
\label{eqn:S3} 
\end{eqnarray}
where $C_3\simeq1$ and $B_3\simeq 1$ 
(see Gazta\~naga 1994 for further details), and $r_3$
can be expressed as
\begin{eqnarray}
r_3=\frac{I_1I_3}{I^2_2},
\label{eqn:r3} 
\end{eqnarray}
with
\begin{eqnarray}
I_k=\int_0^{\infty}x^2dx\Phi^kx^{(3-\gamma)(k-1)}(1+z)^{(3+\epsilon-\gamma)(1-k)},
\label{eqn:I_k} 
\end{eqnarray}
$\Phi(x)$ being defined by equation \ref{eqn:Ndense}.

\subsection{The redshift distribution $N(z)$}

To obtain both the spatial correlation function
and the spatial skewness from the respective projected quantities, we
need $N(z)$, the redshift distribution for
radio sources. 
\begin{figure}
\vspace{8cm}  
\includegraphics{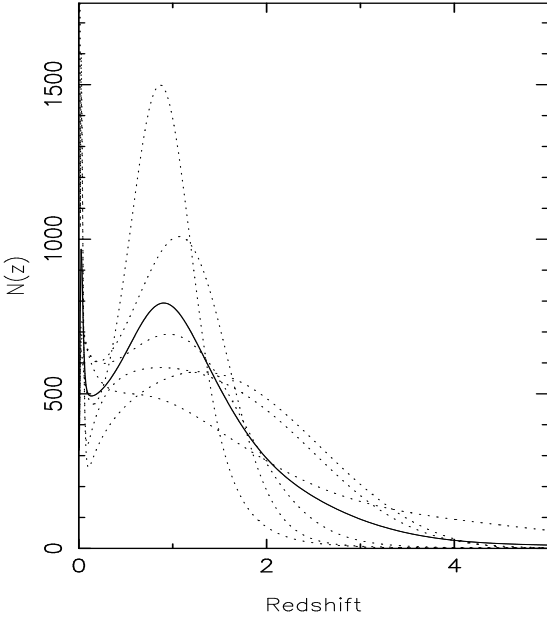} 
\caption{Redshift distribution N(z) for the radio source population at
  1.4 GHz at a flux limit of 3 mJy. The dotted curves represent the 6
  models (1-4, 6-7) taken from Dunlop \& Peacock (1990); the solid curve
  is the average.
\label{fig:Nz_1} }
\end {figure}
\begin{figure}
\vspace{8cm}  
\includegraphics{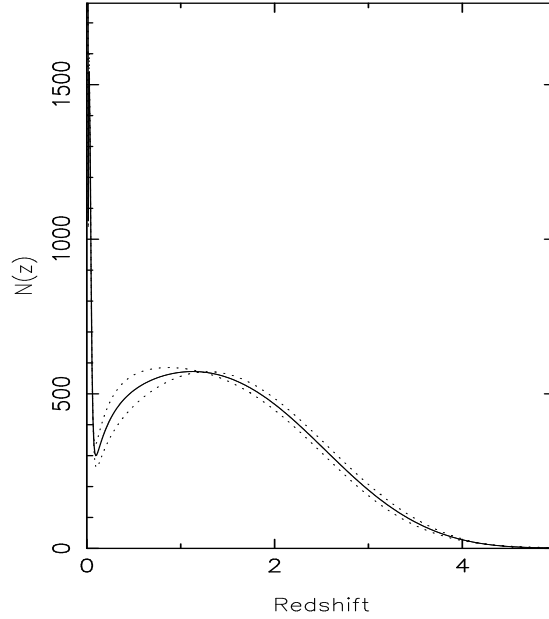} 
\caption{Redshift distribution N(z) for the radio source population at
  1.4 GHz at a flux limit of 3 mJy. The dotted lines are models 6 and
  7 from Dunlop \& Peacock (1990); the solid line is
  the average.
\label{fig:Nz_2} }
\end {figure}
There is not even a completely-identified sample of radio
sources at 3~mJy, leave alone one with measured redshifts.
However, the Dunlop \&
Peacock (DP, 1990) models of epoch-dependent luminosity functions for 
radio sources    can be used to make estimates of $N(z)$. These models 
were derived with Maximum Entropy analysis to determine the coefficients 
for polynomial expansions to represent the epoch-dependent luminosity 
function; the approach incorporated the then-available identification 
and redshift data for complete samples from radio surveys at several 
frequencies. From this we have adopted three different models 
of $N(z)$ for subsequent deprojection analysis.

The first model ($N_1(z)$) is shown in Figure~\ref{fig:Nz_1}. 
The dotted lines represent the six models (1-4, 6-7) taken from DP and
the solid line, the average, is our $N_1(z)$. 
The large spread indicates the uncertainty
due to incomplete or statistically-limited redshift
data as well as the extrapolation of the data to such a low flux density. 
Nevertheless there are
two interesting features which are general. The spike
seen at small redshifts indicates that at such low flux densities, 
the lowest-power tail of the local radio luminosity function begins 
to contribute substantial numbers of low-redhift sources. This spike 
is almost certainly underestimated; the DP analysis did not encompass 
the evolving starburst-galaxy population (e.g. Windhorst et al., 
1985) now believed to constitute a majority of sources at mJy levels 
(see Wall, 1994 for an overview).  The second feature is the  prominent 
maximum around $z\sim 1$ displayed by all models, confirming that the median 
redshift for 
radio sources in radio surveys over a wide flux-density range is a factor 
of 10 higher than that of wide-field optical surveys.

The second model ($N_2(z)$) considered for our deprojections is
represented by the solid line in Figure~\ref{fig:Nz_2} and is obtained  
by averaging models 6 and 7 from DP, models of pure luminosity evolution. 
The more prominent spike at low redshifts may be a more realistic 
representation of the total population.

The third model ($N_3(z)$) is simply 
from $N_1(z)$ with the small-$z$ spike patched out.

We take $N_1(z)$ as our best estimate, and use $N_2(z)$ and $N_3(z)$ to
test how sensitive the deprojection is to the form of $N(z)$. In
particular we wish to see how the results for the correlation length $r_0$
and the quantity $r_3$ in equation \ref{eqn:r3} are affected by 
``extreme'' models (one dominated by the spike and one completely without
it).

\subsection{The Predicted Angular Correlation Function}

The values obtained for the slope $\gamma$ in Section 3.2 are high
compared to the canonical value of 1.7 measured for optically-selected 
galaxies. 
This may be due to the intrinsic clustering properties of radio
galaxies, but in this section we consider another possibility. 
The depth of the survey, $z\sim 1$, means that the angular
scales $0.3^\circ < \theta < 3^\circ$, where we measure clustering,
correspond to linear scales large enough such that the correlation function
$\xi(r,z)$ is no longer a simple power-law, but has steepened from that 
given by $\gamma\simeq 1.7$.

\begin{figure}
\vspace{8cm}  
\includegraphics{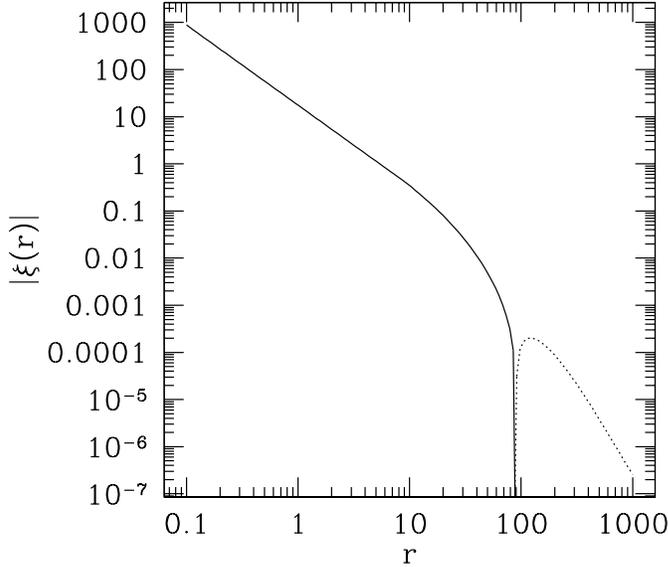} 
\caption{CDM prediction for the correlation function with 
$\Gamma=0.2$. The $y$ axis is the absolute value of $\xi(r)$. 
For $r\sim 100 h^{-1}$ Mpc, 
$\xi(r)$ becomes negative, and we plot its absolute value as the dotted line. }

\label{fig:cdm_xi} 
\end {figure}
\begin{figure}
\vspace{8cm}  
\includegraphics{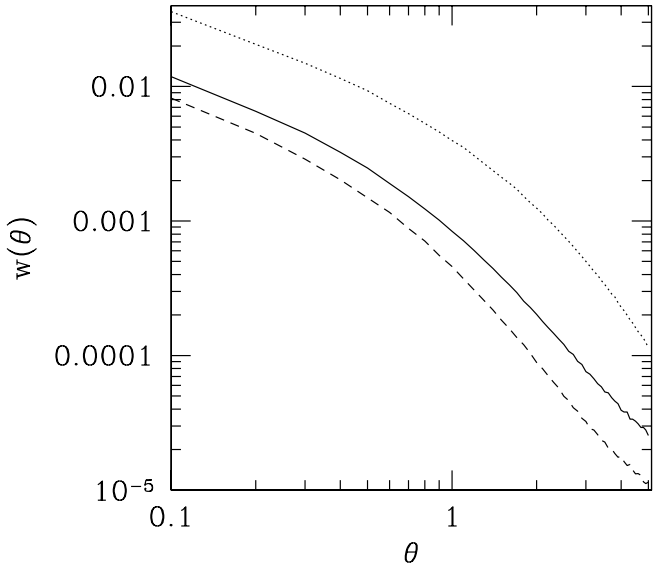} 
\caption{CDM predictions for $w(\theta)$, showing the effects of
  varying the cosmological parameters, with $N(z)\equiv N_1(z)$ as 
described in
  Section 4.3. The dotted line is for $\Omega=1$,
  $h=1$, the solid line for $\Omega=0.3$, $h=0.65$, and the
  dashed line for $\Omega=0.5$, $h=0.5$.
\label{fig:cdm_w} }
\end {figure}

If this is the case, choice of which value of $\gamma$ to use in the
deprojections 
of equations~\ref{eqn:A_power} and \ref{eqn:S3} is not
straightforward.  We have measured the slope $\gamma=2.50$ at large
scales (Section 3.2), but simply using the same slope for all scales
would lead to an overestimate of $w(\theta)$ and therefore of
$\xi(r,z)$ (through equation~\ref{eqn:limber}) at small scales.
Both $r_0$ and $r_3$ are very sensitive to $\gamma$ (see
below) and therefore the assessment of its proper value is very
important for our conclusions.

To investigate what values are expected for $\gamma$ we have
taken a theoretical correlation function $\xi(r,z)$ up to scales
$r\sim 1000 h^{-1} $ Mpc and projected it to get $w(\theta)$ by
directly integrating equation~\ref{eqn:limber}. 
By doing the integrations numerically we do not make the assumption
that $\xi$ is a power law, and do not use any of the approximations
used in Section 4.1. 
In Figure~\ref{fig:cdm_xi} we plot the $\xi(r)$ used.
On large scales it is the linear prediction for a CDM model with
$\Gamma=0.2$.
On small scales the linear-theory prediction underestimates the true
amplitude, and as a rough approximation to allow for this, we simply
extrapolated a power-law of slope $-1.7$ for scales $r<10 h^{-1}$ Mpc, as 
observed from the APM correlation function (Maddox et al., 1990).  
Given the simple heuristic aims of this section, and the large
uncertainties in the evolution of galaxy clustering and bias, more
sophisticated methods to account for non-linearity (e.g. Peacock \&
Dodds, 1996) are unnecessary, though we intend a more careful study 
in a future paper. 
The overall normalization is set so that $r_0 = 5.4 h^{-1}$ Mpc.  

Given this form for $\xi(r)$, the shape and the amplitude of
$w(\theta)$ still depend on the functional
form of $N(z)$ and the values of $\Omega$, $h$. 
Figure~\ref{fig:cdm_w} shows $w(\theta)$ obtained using the redshift
distribution $N_1(z)$ as introduced in Section 4.3, with three
different sets of cosmological parameters.
The curves are significantly different, but they all show the same
power-law behaviour for small angles, with a slope of 
$\sim -0.7$, corresponding to the expected $(1-\gamma)$ with
$\gamma=1.7$;  
then, at $\theta\sim 0.3^\circ$, there is a break from this power law
towards steeper slopes, 
corresponding to $\gamma \sim 3.5 $ for $\theta  \simgt 2^\circ$. 

In the interval $0.3^\circ \la\theta\la 3^\circ$ the curves are close
to a power-law with slope $\simeq -1.5$.
This is in agreement with the value of $\gamma \approx 2.5$ obtained
in Section 3.2.
The smaller the value of $\Omega$, the steeper is the effective slope. 
Of the three curves in the range
$0.3^\circ\la\theta\la 3^\circ$, that for $\Omega=0.3$ and
$h=0.65$ fits the data best. 
However this rather superficial analysis assumes a
fixed $\xi(r)$ and $N(z)$, both of which will significantly change the
shape of $w$.
We intend to make a more thorough comparison between models and the
data in a future paper.

\subsection {Results}

We now estimate the correlation length $r_0$ and the spatial 
normalized skewness
$S_3^*(R)$ using equations~\ref{eqn:A_power}, \ref{eqn:S3}, \ref{eqn:r3}
and \ref{eqn:I_k} with different redshift distributions, and a range
of values for $\gamma$ and $\epsilon$.
We confine attention to our most reliable clustering estimates,
namely those
obtained in Section 3.2 from the C3 catalogue. 

\begin{figure*}
\vspace{15cm}  
\includegraphics{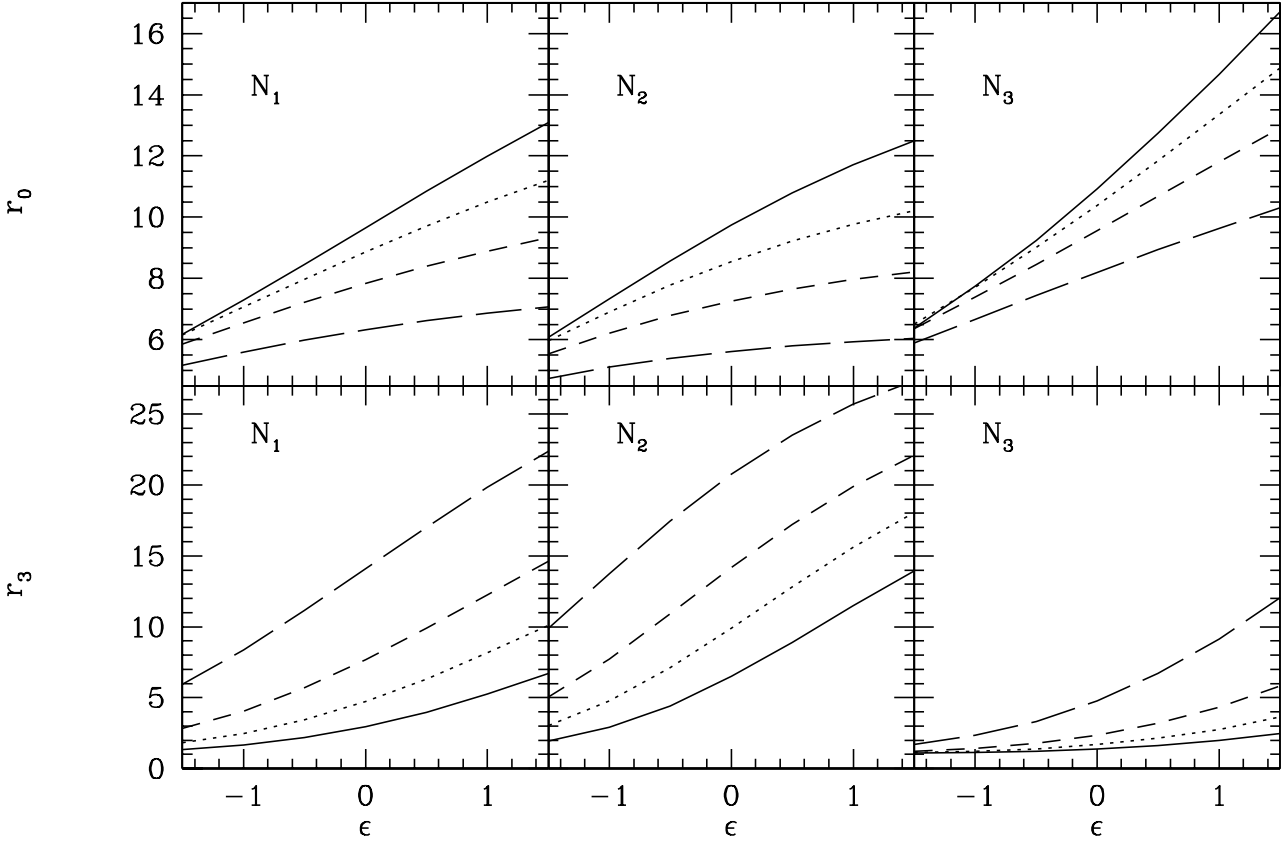} 
\caption{Correlation length $r_0$ and $r_3\sim S_3(\Theta)/S_3^*(R)$
  as a function of the evolution
  parameter $\epsilon$ for different values of $\gamma$ and
  respectively  
  $N_1$, $N_2$, $N_3$, in the case of the C3 catalogue; 
the solid line corresponds to to 
 $\gamma=1.8$, the dotted line to $\gamma=2.0$, the
  short-dashed line to $\gamma=2.2$, and the long-dashed line
  to $\gamma=2.50$.}
\label{fig:r0_epsilon1}
\end{figure*}

Figure~\ref{fig:r0_epsilon1} shows the trends
for both the correlation length $r_0$ and the quantity $r_3$, as
functions of the evolution parameter $\epsilon$, obtained for
different values of $\gamma$ and the three different models for the
redshift distribution $N(z)$ introduced in Section 4.3.
If we fix a value for $\gamma$ and vary the form of $N(z)$, the
variations for $r_0$ ($\la$ 10\%) are not as dramatic as those for
$r_3$. In general, the stronger the low-redshift spike in N(z) the
smaller the value obtained for the correlation length $r_0$.
If instead we fix the functional form for $N(z)$, varying $\gamma$
strongly affects both $r_0$ and $r_3$.
Note also that the value of the amplitude $A$ used in the deprojection
equation~\ref{eqn:A_power} to obtain the correlation length $r_0$ also
depends on $\gamma$ through the quantity $C_{\gamma}$
(equation~\ref{eqn:sigma2}).

Armed with these results we face choosing the value for $\gamma$ to 
provide the best estimates for $r_0$ and $r_3$.
Our counts-in-cells analysis carried out for scales
$\Theta\le0.3^\circ$ is dominated by the shot (Poisson) noise, so we
must consider clustering on larger scales.
For $0.3^\circ \le \Theta\le 3^\circ$ we observe a slope $\gamma=2.50$
but from Section 4.4 it follows that clustering on these angular
scales is determined by the spatial correlation function on scales
larger than $10 h^{-1} $ Mpc where a power law is not a good approximation.
As mentioned in section 4.4, using $\gamma=2.50$ in
equation~\ref{eqn:w_power} will overestimate $w(\theta)$, and hence
$\xi(r)$ at small scales where the spatial correlation function is
noticeably greater than 1 (see Figure~\ref{fig:cdm_xi}).
This will severely bias the value of $r_0$ obtained from integrating
equation~\ref{eqn:A_power}.
This problem is even worse when we try to deproject the skewness, for
the quantity $r_3$ appearing in equation~\ref{eqn:r3} depends on the
square of the correlation function (see Section 3.3).
The choice of $\gamma$ is further complicated by the fact that $N(z)$
suggests the presence of at least two populations of radio
sources, which may
well have different clustering properties.
Yet another uncertainty is introduced by fact that $\gamma$ may well be a
function of redshift, even for a single population.

For convenience of comparison with other studies we quote our
estimates using the ``standard value''  $\gamma =1.8$ and assume 
stable clustering ($\epsilon=0$). 
Adopting $S_3(\Theta)=12.5$, we then find 
\begin{eqnarray}
N_1(z):\;\;\;\;\;
r_0\sim 9.7 h^{-1}Mpc;\;\;S_3^*(R)\simeq\frac{S_3(\Theta)}{r_3}\sim
4.2, \nonumber\\
N_2(z):\;\;\;\;\;
r_0\sim 9.7 h^{-1}Mpc;\;\;S_3^*(R)\simeq\frac{S_3(\Theta)}{r_3}\sim
1.2 ,\nonumber\\
N_3(z):\;\;\;\;\;
r_0\sim 10.9 h^{-1}Mpc;\;\;S_3^*(R)\simeq\frac{S_3(\Theta)}{r_3}\sim
8.9 .\nonumber\\
\label{eqn:r0_S3} 
\end {eqnarray}
We note that the values for $r_0$ are little affected by the 
choice of $N(z)$. 
Our estimate of $r_0$ for radio sources is
larger than the value for optical  
($r_0 \sim 5 h^{-1} $ Mpc) and IRAS galaxies ($r_0\sim 4 h{-1}$ Mpc) 
and smaller than 
for the cluster-cluster correlation length 
($r_0 \sim 14-20  h^{-1} $ Mpc).  

On the other hand, we see that the spatial normalized skewness 
$S_3^*(R)$  is sensitive to $N(z)$.
(We have neglected the
small corrections for the difference between square and circular
cells.)
Apart from the observational uncertainties 
in estimating $S_3^*(R)$ 
for the radio sources, we should remember that comparison with the 
theoretical value of $S_3^*(R)$ for the {\it mass} perturbations 
(equation~\ref{eqn:s_3star}) depends on how radio galaxies are biased
relative to the mass distribution. 
This bias may also be epoch-dependent 
(e.g. Fry, 1996). 
Nevertheless, our estimates are in accord with the prediction 
(equation~\ref{eqn:s_3star}) of $S_3^*(R) \sim 2.9$ 
for  power-spectrum $P(k) \sim k^{-1}$ 
expected (based on other observations and 
models) on the weakly non-linear scales $R \sim 5-15 h^{-1} $ Mpc 
probed by our analysis. (Note that angular scale 
of 1 degree at the median redshift 
of the survey, ${\bar z} \sim 1$, corresponds to comoving distance
of $30 h^{-1}$ Mpc for an Einstein - de Sitter universe.)
It is also interesting to note that our values for $S_3^*$ bracket 
the values derived from the optical APM survey
($S_3^* \sim 3.2$; Gazta\~naga et al., 1994) and from the IRAS survey 
($S_3^* \sim 2.8$; Kim \& Strauss, 1998), even though the observed
difference in the spatial skewness could reflect differences in
clustering properties between radio sources (possible mix of more
objects like bright ellipticals and starbursting galaxies) and
optical or IRAS galaxies.

\section {CONCLUSIONS}

By investigating the distribution function (counts in cells) 
for radio sources of the FIRST
survey, we have shown how to infer the 
clustering properties of host radio galaxies. 
We considered three different catalogues above 3~mJy, generated
from the original survey by following different procedures for
combining source components. Focusing on the catalogue 
obtained with 
the most sensible procedure, by relating the second moment of the
distribution to the angular two-point correlation function, we find
a power-law behaviour for $w(\theta)$.

From the analysis of the third moment we have shown how variance
and skewness are related through the functional form
$\Psi_3=S_3(\Psi_2)^{\alpha}$ with $\alpha\simeq 2$ 
and $S_3=const$ with angular scale. 
By inverting the projected quantities we have estimated
the spatial correlation length $r_0 \sim 10 h^{-1}$Mpc
and the spatial skewness $S^*_3(R)\sim 1-9 $.
While the value for $r_0$ is relatively independent of the
functional form of the redshift distribution $N(z)$, the value
for skewness is strongly dependent on it.

Our results indicate 
that the large-scale clustering of radio sources is in
accord with the gravitational instability picture for the growth of
perturbations from a primordial Gaussian field.
Our measurements of deviations from a Gaussian distribution 
do not seem to require initial non-Gaussian perturbations 
(e.g. cosmic strings or texture).

There are crucial observations required to further this analysis. We need
a better understanding of populations and source structures at mJy levels
to remove statistical uncertainties due to multi-component sources.  Even
more important is the observational measurement of N(z), from
identifications and redshift measurements for a complete sample. Radio
morphologies and optical studies for a small flux-limited sample would
achieve both goals.

\vspace{1cm}
\noindent
{\bf ACKNOWLEDGEMENTS}\\
MM acknowledges support from the Isaac Newton Scholarship. We thank
Catherine Cress, George
Efstathiou and David Helfand for helpful discussions.


\begin{thebibliography}{}

\bibitem[\protect\citename{Baleisis et al. }1998]{Blow}
Baleisis A., Lahav O., Loan A.J. \& Wall J.V., 1998; {\it MNRAS}, in press
\bibitem[\protect\citename{Becker et al. }1995]{Be}
Becker R.H., White R.L., Helfand D.J., 1995; {\it ApJ}, {\bf 450}, 559
\bibitem[\protect\citename{Benn \& Wall }1995]{Benn}
Benn C.R., Wall J.V., 1995; {\it MNRAS}, {\bf 272}, 678
\bibitem[\protect\citename{Coles et al. }1991]{Co}
Coles P., Frenk C.S., 1991; {\it MNRAS}, {\bf 253}, 727
\bibitem[\protect\citename{Cress et al. }1996]{Cr}
Cress C.M., Helfand D.J., Becker R.H., Gregg M.D., White R.L., 1996; 
{\it ApJ}, {\bf 473}, 7
\bibitem[\protect\citename{Dunlop J.S., Peacock J.A.}1990]{Du}
Dunlop J.S., Peacock J.A., 1990; {\it MNRAS}, {\bf 247}, 19
\bibitem[\protect\citename{Fisher et al. }1993]{Fi}
Fisher K.B., Davis M., Strauss M.A., Yahil A., Huchra J.P., 1993; 
{\it ApJ}, {\bf 402}, 42
\bibitem[\protect\citename{Frieman \& Gazta\~naga }1994]{Fri}
Frieman J.A., Gazta\~naga E., 1994; {\it ApJ}, {\bf 425}, 392
\bibitem[\protect\citename{Fry }1996]{Fry}
Fry J., 1996; {\it ApJ}, {\bf 461}, L65
\bibitem[\protect\citename{Fry \& Gazta\~naga }1993]{Fr}
Fry J.N., Gazta\~naga E., 1993; {\it ApJ}, {\bf413}, 447
\bibitem[\protect\citename{Gazta\~naga }1994]{Gaz}
Gazta\~naga E., 1994; {\it MNRAS}, {\bf 268}, 913
\bibitem[\protect\citename{Gazta\~naga }1995]{Ga}
Gazta\~naga E., 1995; {\it ApJ}, {\bf 454}, 561
\bibitem[\protect\citename{Gazta\~naga \& Bernardeau }1997]{GaB}
Gazta\~naga E., Bernardeau F., 1997; Astro-ph/9707095
\bibitem[\protect\citename{Gregory }1991]{Gre}
Gregory P.C., Condon J.J., 1991; {\it ApJS}, {\bf 75}, 1011
\bibitem[\protect\citename{Griffith }1993]{Gri}
Griffith M.R., Wright A.E., 1993; {\it AJ}, {\bf 105}, 1666
\bibitem[\protect\citename{Hamilton }1993]{Ha}
Hamilton A.J., 1993; {\it ApJ}, {\bf 417}, 19
\bibitem[\protect\citename{Juszkiewicz et al. }1995]{Jus}
Juszkiewicz R., Weinberg D.H., Amsterdamski P., Chodorowski M., 
Bouchet F.R., 1995; {\it ApJ}, {\bf 42}, 39 
\bibitem[\protect\citename {Kaiser }1984]{Ka}
Kaiser N., 1984; {\it ApJ}, {\bf 284}, L9
\bibitem[\protect\citename{Kim \& Strauss}1998]{ks98}
Kim R.S.J., Strauss M.A., 1998; {\it ApJ}, {\bf 493}, 39 
\bibitem[\protect\citename{Kooiman et al. }1995]{Ko}
Kooiman L.K., Burns J.O., Klypin A.A., 1995; {\it ApJ}, {\bf 448}, 500
\bibitem[\protect\citename{Lahav et al. }1992]{La}
Lahav O., Saslaw W.C., 1992; {\it ApJ}, {\bf 396}, 430
\bibitem[\protect\citename{Landy \& Szalay}1993]{Lasy}
Landy S.D., Szalay A.S., 1993; {\it ApJ}, {\bf 412}, 64
\bibitem[\protect\citename{Limber }1953]{Li}
Limber D.N., 1953; {\it ApJ}, {\bf 117}, 134
\bibitem[\protect\citename{Loan et al. }1997]{Lo}
Loan A.J., Wall J.V., Lahav O., 1997; {\it MNRAS}, {\bf 286}, 994
\bibitem[\protect\citename{Maddox et al. }1990]{Ma}
Maddox S.J., Efstathiou G., Sutherland W.J., Loveday J., 1990;
{\it MNRAS}, {\bf242}, 43P
\bibitem[\protect\citename{Mo et al. }1992]{Mo}
Mo H.J., Jing Y.P., Borner G., 1992;
{\it ApJ}, {\bf392}, 452
\bibitem[\protect\citename{Oort } 1987]{Oo}
Oort M.J.A., 1987; {\it PhD Thesis}, Leiden Observatory
\bibitem[\protect\citename{Peacock and Dodds}1996]{PD}
Peacock J.A., Dodds S.J., 1996; {\it MNRAS}, {\bf 280}, L19
\bibitem[\protect\citename{Peebles }1980]{Pe}
Peebles P.J.E., 1980; {\it The Large-Scale Structure of the Universe},
Princeton University Press
\bibitem[\protect\citename{Seldner }1981]{Se}
Seldner M., Peebles P.J.E., 1981; {\it MNRAS}, {\bf 194}, 251
\bibitem[\protect\citename{Shaver et al. }1989]{Sha}
Shaver P.A., Pierre M., 1989; {\it A\&A}, {\bf 220}, 35
\bibitem[\protect\citename{Sicotte }1995]{Si}
Sicotte H, 1995; {\it PhD thesis}, Princeton University
\bibitem[\protect\citemname{Totsuji et al.}1969]{To}
Totsuji H., Kihara T., 1969; {\it PASJ}, {\bf 21}, 221
\bibitem[\protect\citemname{Treyer M.A., Lahav, O.}1996]{Tr}
Treyer M.A., Lahav O., 1996; {\it MNRAS}, {\bf 280}, 469
\bibitem[\protect\citename{Wall}1994]{Wa} 
Wall J.V., 1994; {\it Austr. J. Phys.}, {\bf 47}, 625
\bibitem[\protect\citename{Webster }1976]{We}
Webster A., 1976; {\it MNRAS}, {\bf 175}, 61
\bibitem[\protect\citename{White et al. } 1997]{Whi}
White R.L., Becker R.H., Helfand D.J., Gregg M.D., 1997;
{\it ApJ}, {\bf 475}, 479
\bibitem[\protect\citename{Windhorst et al.}1985]{Wi} 
Windhorst R.A., Miley G.K., Owen F.N., Kron R.G., Koo R.C., 1985; 
{\it ApJ}, {\bf 289}, 494


\end{thebibliography}
\end{document}